\documentclass[twocolumn,american,english]{revtex4}
\usepackage[T1]{fontenc}
\usepackage[latin9]{inputenc}
\usepackage{color}
\usepackage{amsmath}
\usepackage{amssymb}
\usepackage{graphicx}

\makeatletter
\@ifundefined{textcolor}{}
{%
 \definecolor{BLACK}{gray}{0}
 \definecolor{WHITE}{gray}{1}
 \definecolor{RED}{rgb}{1,0,0}
 \definecolor{GREEN}{rgb}{0,1,0}
 \definecolor{BLUE}{rgb}{0,0,1}
 \definecolor{CYAN}{cmyk}{1,0,0,0}
 \definecolor{MAGENTA}{cmyk}{0,1,0,0}
 \definecolor{YELLOW}{cmyk}{0,0,1,0}
}

\makeatother

\usepackage{babel}
\begin{document}
\title{Natural orbital functional for spin-polarized periodic systems}
\author{Raul Quintero-Monsebaiz$^{1,2}$, Ion Mitxelena$^{2,3}$, Mauricio
Rodríguez-Mayorga$^{2}$, Alberto Vela$^{1}$, Mario Piris$^{2,3,4}$}
\email{mario.piris@ehu.eus}

\address{$^{1}$Departamento de Química, Centro de Investigación y de Estudios
Avanzados, Av. Instituto Politécnico Nacional 2508, D. F. 07360, México;}
\address{$^{2}$Donostia International Physics Center (DIPC), 20018 Donostia,
Euskadi, Spain;}
\address{$^{3}$Kimika Fakultatea, Euskal Herriko Unibertsitatea (UPV/EHU),
P.K. 1072, 20080 Donostia, Euskadi, Spain;}
\address{$^{4}$IKERBASQUE, Basque Foundation for Science, 48013 Bilbao, Euskadi,
Spain.\vspace{0.5cm}
}
\begin{abstract}
Natural orbital functional theory is considered for systems with one
or more unpaired electrons. An extension of the Piris natural orbital
functional (PNOF) based on electron pairing approach is presented,
specifically, we extend the independent pair model, PNOF5, and the
interactive pair model PNOF7 to describe spin-uncompensated systems.
An explicit form for the two-electron cumulant of high-spin cases
is only taken into account, so that singly occupied orbitals with
the same spin are solely considered. The rest of electron pairs with
opposite spins remain paired. The reconstructed two-particle reduced
density matrix fulfills certain N-representability necessary conditions,
as well as guarantees the conservation of the total spin. The theory
is applied to model systems with strong non-dynamic (static) electron
correlation, namely, the one-dimensional Hubbard model with periodic
boundary conditions and hydrogen rings. For the latter, PNOF7 compares
well with exact diagonalization results so the model presented here
is able to provide a correct description of the strong-correlation
effects.
\end{abstract}
\maketitle
The ground-state energy of any electronic system could be computed
using the first-order reduced density matrix (1-RDM) according to
Gibert's theorem \citep{Gilbert1975} along with the works of Donnelly
and Parr \citep{Donnelly1978}, Levy \citep{Levy1979} and Valone
\citep{Valone1980}. In most calculations, the spectral decomposition
of the 1-RDM is used to approximate the energy in terms of the naturals
orbitals $\left\{ \phi_{i}\right\} $ and their occupation numbers
$\left\{ n_{i}\right\} $. This formulation is called natural orbital
functional (NOF) theory (NOFT). It is important to note that attaining
the exact energy in density functional theory requires a greater effort
than in NOFT, since the non-interacting part of the electronic Hamiltonian
is actually a one-particle operator, so it has an explicit dependence
on the 1-RDM. A detailed account of the state of the art of the NOFT
can be found elsewhere \citep{Piris2007,Piris2014a,Pernal2016}.

So far, the exact reconstruction of the functional has not been achieved,
therefore, we have to settle for making approximations. The most practical
approach is to reconstruct solely the two-particle RDM (2-RDM) and
employ the well-known energy expression in terms of it. Approximating
the energy in that way implies that the functional still depends on
the 2-RDM \citep{Donnelly1979}, hence the functional $N$-representability
problem arises \citep{Ludena2013,Piris2018}. We have to impose the
$N$-representability conditions \citep{Mazziotti2012a} to the 2-RDM
reconstructed in terms of 1-RDM to guarantee that our approximate
ground-state energy has physical sense. As far as we know, only the
functionals developed by Piris and collaborators (PNOFi, i=1-7) \citep{Piris2013b,Piris2014c,Piris2017}
rely on the reconstruction of the 2-RDM subject to necessary $N$-representability
conditions.

Within NOFT, few attempts have been made to describe systems with
spin polarization. The first calculations in this type of systems
were probably reported in \citep{Goedecker1998} by Goedecker and
Umrigar, who used a NOF that can be traced back to Müller (M) \citep{Muller1984}
and Buijse and Baerends (BB) \citep{Buijse1991} proposals. A formulation
of this functional considering spin-independent natural orbitals (NOs)
but spin-dependent occupation numbers (ONs) was applied by Lathiotakis
et al. \citep{Lathiotakis2005} to the first-row atoms. The major
shortcoming of this approach is that it does not conserve, in general,
the total spin. Rohr and Penal \citep{Rohr2011} showed that MBB functional
has a non-positive fractional spin error, with the exception of one-electron
systems. In fact, extensive violations of $N$-representability conditions
for this functional, and others related to it, have been reported
\citep{RodriguezMayorga2017}.

The discovery \citep{Klyachko2006,Altunbulak2008} of a systematic
way to derive pure-state $N$-representability conditions for the
1-RDM, also known as generalized Pauli constraints (GPCs), has lately
allowed to gain new insights on spin-uncompensated systems \citep{Theophilou2015},
as well as to open a new way to develop NOFs \citep{Benavides-Riveros2018}.
Indeed, necessary pure $N$-representability conditions on the 2-RDM
can be derived by applying the GPCs to a family of effective 1-RDMs
\citep{Mazziotti2016}. It should be noted that the application of
GPCs restricts the 1-RDM variational space leading to improvements
in energy, but it does not improve the reconstruction of the approximate
functional per se. A 1-RDM that represents a pure state does not guarantee
that the reconstructed electron-electron potential energy will be
pure-state $N$-representable, except if the reconstruction is the
exact one. The functional $N$-representability problem continues
to exist for pure 1-RDMs when we make approximations for the functional,
and it is still related to the $N$-representability of the 2-RDM. 

The first extension of the Piris reconstruction functional to high-spin
multiplet states was reported in the past for PNOF1 \citep{Leiva2007}.
Later, a necessary condition to ensure the conservation of the total
spin was obtained for the two-particle cumulant matrix \citep{Piris2009}.
PNOF3 showed exceptional performance for atoms and molecules without
spin compensation \citep{Piris2010}. Unfortunately, closer analysis
of the dissociation curves for various diatomics revealed that PNOF3
overestimates the amount of electron correlation, when the non-dynamic
electron correlation becomes important. It was demonstrated that this
ill behavior was related to the violation of the $N$-representability
conditions, in particular, to the violation of the G positivity conditions
for the 2-RDM \citep{Piris2010a}. 

In this work, we focus on two recent approximations that are based
on pairs of electrons with opposite spins \citep{Piris2018a}. The
first is the independent pair model PNOF5 \citep{Piris2011,Piris2013e},
which remarkably turned out to be strictly pure $N$-representable
\citep{Pernal2013,Piris2013c}. PNOF5 takes into account the important
part of the dynamical electron correlation corresponding to the intra-pair
interactions, and most of the non-dynamical effects. However, no inter
pair electron correlation is accounted. To include the missing interactions,
PNOF6 \citep{Piris2014c} and PNOF7 \citep{Piris2017} were proposed.
The latter was recently \citep{mitxelena2018a} improved by an adequate
choice of sign factors for the inter-pair interactions. Since PNOF7
provides a robust description of non-dynamic correlation effects,
we will limit ourselves to this interacting pair model.

Our goal is to extend PNOF5 and PNOF7 to describe systems with spin
polarization. We follow the spin-restricted formulation proposed in
\citep{Leiva2007} which has the virtue of avoiding spin contamination
effects. Our hypothesis is to take the occupation numbers of singly
occupied orbitals that contribute to the spin of the system equal
to one, whereas the rest of electron pairs with opposite spins are
distributed in obital subspaces, with one pair per subspace. In the
latter, the orbital occupancies are fractional, which account for
the electron correlation. The resulting functionals are tested in
energy calculations for model systems with strong non-dynamic electron
correlation, namely, the one-dimensional Hubbard model with periodic
boundary conditions and hydrogen rings up to 16 atoms, in order to
compare with exact diagonalization values.

\section{Theory}

In NOFT, the electronic energy is given in terms of the NOs and their
ONs, namely,
\begin{equation}
E=\sum\limits _{i}n_{i}\mathcal{H}_{ii}+\sum\limits _{ijkl}D[n_{i},n_{j},n_{k},n_{l}]<kl|ij>\label{NOF}
\end{equation}

where $\mathcal{H}_{ii}$ denotes the diagonal elements of the one-particle
part of the Hamiltonian involving the kinetic energy and the external
potential operators, $<kl|ij>$ are the matrix elements of the two-particle
interaction, and $D[n_{i},n_{j},n_{k},n_{l}]$ represents the reconstructed
2-RDM from the ONs. Restriction of the ONs to the range $0\leq n_{i}\leq1$
represents a necessary and sufficient condition for ensemble $N$-representability
of the 1-RDM \citep{Coleman1963} under the normalization condition
$\sum_{i}n_{i}=N$.

The $N$-electron Hamiltonian commonly used in electronic calculations
does not contain any spin coordinates. Consequently, the eigenfunctions
of the Hamiltonian are also eigenfunctions of both spin operators
$\hat{S}_{z}$ and $\hat{S}^{2}$. For $\hat{S}_{z}$ eigenstates,
only density matrix blocks that conserve the number of each spin type
are non-vanishing. Specifically, the 1-RDM has two nonzero blocks
$\alpha$ and $\beta$, whereas the 2-RDM has three independent nonzero
blocks: $D^{\alpha\alpha}$, $D^{\alpha\beta}$, and $D^{\beta\beta}$. 

Let us divide the spin-orbital set $\left\{ \phi_{i}\left(\mathbf{x}\right)\right\} $
into two subsets: $\left\{ \varphi_{p}^{\alpha}\left(\mathbf{r}\right)\alpha\left(\mathbf{s}\right)\right\} $
and $\left\{ \varphi_{p}^{\beta}\left(\mathbf{r}\right)\beta\left(\mathbf{s}\right)\right\} $.
In order to avoid spin contamination effects, the spin-restricted
theory is employed, in which a single set of orbitals is used for
$\alpha$ and $\beta$ spins: $\varphi_{p}^{\alpha}\left(\mathbf{r}\right)=\varphi_{p}^{\beta}\left(\mathbf{r}\right)=\varphi_{p}\left(\mathbf{r}\right)$.
Accordingly, the expectation values of the spin operators read as
\citep{Piris2007}
\begin{align}
<\hat{S}_{z}>= & \frac{N^{\alpha}-N^{\beta}}{2}=M_{S}\\
<\hat{S}^{2}>= & \frac{N\left(4-N\right)}{4}{\displaystyle +\sum\limits _{pq}}\left\{ D_{pq,pq}^{\alpha\alpha,\alpha\alpha}+D_{pq,pq}^{\beta\beta,\beta\beta}-2D_{pq,qp}^{\alpha\beta,\alpha\beta}\right\} \nonumber 
\end{align}
The matrix elements of the 2-RDM can conveniently be expressed in
terms of the cumulant expansion: 
\begin{align}
D_{pq,rt}^{\sigma\sigma,\sigma\sigma} & =\frac{n_{p}^{\sigma}n_{q}^{\sigma}}{2}\left(\delta_{pr}\delta_{qt}-\delta_{pt}\delta_{qr}\right)+\lambda_{pq,rt}^{\sigma\sigma,\sigma\sigma}\,_{(\sigma=\alpha,\beta)}\\
D_{pq,rt}^{\alpha\beta,\alpha\beta} & =\frac{n_{p}^{\alpha}n_{q}^{\beta}}{2}\delta_{pr}\delta_{qt}+\lambda_{pq,rt}^{\alpha\beta,\alpha\beta}
\end{align}
where $\mathbf{\lambda}$ is the cumulant matrix \citep{Mazziotti1998}.
Using this expansion, the energy (\ref{NOF}) reads
\begin{align}
E & =\sum\limits _{p}\left(n_{p}^{\alpha}+n_{p}^{\beta}\right)\mathcal{H}_{pp}+\frac{1}{2}\sum\limits _{pq}\left(n_{q}^{\alpha}+n_{q}^{\beta}\right)\left(n_{p}^{\alpha}+n_{p}^{\beta}\right)\mathcal{J}_{pq}\nonumber \\
 & -\frac{1}{2}\sum\limits _{pq}\left(n_{q}^{\alpha}n_{p}^{\alpha}+n_{q}^{\beta}n_{p}^{\beta}\right)\mathcal{K}_{pq}+\sum\limits _{pqrt}\text{ }\widetilde{\lambda}_{pq,rt}\left\langle rt|pq\right\rangle \label{NOFcum}
\end{align}
 where $\mathcal{J}_{pq}=\left\langle pq|pq\right\rangle $ and $\mathcal{K}_{pq}=\left\langle pq|qp\right\rangle $
are the usual Coulomb and exchange integrals, respectively. $\widetilde{\lambda}_{pq,rt}$
denotes the spinless cumulant matrix, 
\begin{equation}
\widetilde{\lambda}_{pq,rt}=\lambda_{pq,rt}^{\alpha\alpha,\alpha\alpha}+\lambda_{pq,rt}^{\alpha\beta,\alpha\beta}+\lambda_{qp,tr}^{\alpha\beta,\alpha\beta}+\lambda_{pq,rt}^{\beta\beta,\beta\beta}\label{cumsl}
\end{equation}

From the sum rules of the 2-RDM blocks, it can be easily shown \citep{Piris2007}
that the expectation value of $\widehat{S}^{2}$ is
\begin{equation}
<\hat{S}^{2}>={\textstyle \frac{N^{\alpha}+N^{\beta}}{2}+\frac{\left(N^{\alpha}-N^{\beta}\right)^{2}}{4}-\sum\limits _{p}n_{p}^{\alpha}n_{p}^{\beta}-2\sum\limits _{pq}\lambda_{pq,qp}^{\alpha\beta,\alpha\beta}}
\end{equation}
For a given value $S$, there are $\left(2S+1\right)$ energy degenerate
spin-multiplet states: $|SM_{S}>$, $M_{S}=-S,...,S$. Let us focus
on the high-spin multiplet state ($M_{S}=S$) and assume that $n_{p}^{\alpha}=n_{p}+m_{p\,},\;n_{p}^{\beta}=n_{p}$
so that $n_{p}$ and $m_{p}$ fulfill the following constrains: 
\begin{equation}
\sum\limits _{p}m_{p}=2S\,,\;2\sum\limits _{p}n_{p}=N-2S\label{npmp}
\end{equation}
Then, $<\widehat{S}^{2}>=S(S+1)$ implies \citep{Piris2009} the following
sum rule for the cumulant $\alpha\beta$-block:
\begin{equation}
2\sum\limits _{q}\lambda_{pq,qp}^{\alpha\beta,\alpha\beta}=2\sum\limits _{q}\lambda_{qp,pq}^{\alpha\beta,\alpha\beta}=n_{p}-n_{p}^{\alpha}n_{p}^{\beta}\label{cumab}
\end{equation}
In general, the cumulant $\lambda$ has a dependence of four indexes,
so it is expensive from the computational point of view to use such
quantities. We shall use the reconstruction functional proposed in
Ref. \citep{Piris2006}, in which the two-particle cumulant is explicitly
constructed in terms of two-index matrices, $\Delta$ and $\Pi$.
The latter are selected to satisfy necessary $N$-representability
conditions and sum rules by the 2-RDM. A systematic application of
$N$-representability conditions in the reconstruction of $\lambda$
has led to the PNOF series \citep{Piris2013b,Piris2014c,Piris2017}.
The latter has the following spin structure:
\begin{align}
\lambda_{pq,rt}^{\sigma\sigma,\sigma\sigma} & =-\frac{\Delta_{pq}^{\sigma\sigma}}{2}\left(\delta_{pr}\delta_{qt}-\delta_{pt}\delta_{qr}\right)\:_{(\sigma=\alpha,\beta)}\\
\lambda_{pq,rt}^{\alpha\beta,\alpha\beta} & =-\frac{\Delta_{pq}^{\alpha\beta}}{2}\delta_{pr}\delta_{qt}+\frac{\Pi_{pr}}{2}\delta_{pq}\delta_{rt}\label{pnof_ab}
\end{align}
which leads to the energy functional
\begin{align}
E & =\sum\limits _{p}\left(n_{p}^{\alpha}+n_{p}^{\beta}\right)\mathcal{H}_{pp}+\sum\limits _{pr}\Pi_{pr}\mathcal{L}_{rp}\nonumber \\
 & +\frac{1}{2}\sum\limits _{pq}\left[\left(n_{q}^{\alpha}+n_{q}^{\beta}\right)\left(n_{p}^{\alpha}+n_{p}^{\beta}\right)-\widetilde{\Delta}_{pq}\right]\mathcal{J}_{pq}\nonumber \\
 & -\frac{1}{2}\sum\limits _{pq}\left[\left(n_{q}^{\alpha}n_{p}^{\alpha}+n_{q}^{\beta}n_{p}^{\beta}\right)-\left(\Delta_{pq}^{\alpha\alpha}+\Delta_{pq}^{\beta\beta}\right)\right]\mathcal{K}_{pq}\label{PNOF}
\end{align}

Here, $\Delta^{\sigma\sigma'}$ are real symmetric matrices, whereas
$\Pi$ is a spin-independent Hermitian matrix. These matrices satisfy
several constraints imposed by the sum rules and $N$-representability
conditions of the two-particle cumulant. $\widetilde{\Delta}_{pq}$
denotes the spinless $\Delta$ matrix, 
\begin{equation}
\widetilde{\Delta}_{pq}=\Delta_{pq}^{\alpha\alpha}+\Delta_{pq}^{\alpha\beta}+\Delta_{qp}^{\alpha\beta}+\Delta_{pq}^{\beta\beta}
\end{equation}
whereas the new integral $\mathcal{L}_{rp}=\left\langle rr|pp\right\rangle $
arises from the correlation between particles with opposite spins
and is called the exchange and time-inversion integral \citep{Piris1999}.

By combining the sum rule (\ref{cumab}) with the ansatz (\ref{pnof_ab}),
one arrives at the following diagonal elements \citep{Piris2009}
\begin{equation}
\Delta_{pp}^{\alpha\beta}=n_{p}^{\alpha}n_{p}^{\beta}\,,\;\Pi_{pp}=n_{p}
\end{equation}
that guarantee the conservation of the total spin.

In this work, we restrict ourselves to pairing-based approximations
PNOF5 and PNOF7, that have been successfully implemented for singlets.
Here, we extend both NOFs to situations in which a system could have
one or more unpaired electrons, so we can have spin-polarized systems. 

Consider $N_{U}$ unpaired electrons, and $N_{P}$ paired electrons,
so that $N_{U}+N_{P}=N$. Similarly, divide the orbital space $\Omega$
into two subspaces: $\Omega=\Omega_{U}\oplus\Omega_{P}$. In $\Omega_{P}$,
the spatial orbitals are double occupied ($n_{p}\neq0$), whereas
singly occupied orbitals ($n_{p}=0$) can only be found in $\Omega_{U}$.
We assume further that all spins corresponding to the $N_{P}$ electrons
are coupled as a singlet, so the occupancies for particles with $\alpha$
and $\beta$ spin are equal:
\begin{equation}
n_{p}^{\alpha}=n_{p}^{\beta}=n_{p}\,,\quad m_{p}=0\,,\quad p\in\Omega_{P}
\end{equation}

According to the electron-pairing approach in NOFT \citep{Piris2018a},
the orbital space $\Omega_{P}$ is in turn divided into $N_{P}/2$
mutually disjoint subspaces $\Omega_{P}=\Omega_{1}\oplus\Omega_{2}\oplus...\oplus\Omega_{N_{P}/2}$.
Each subspace $\Omega{}_{g}$ contains one orbital $g$ below the
level $N_{P}/2$, and $N_{g}$ orbitals above it. In what follows,
we consider $N_{g}$ equal to a fixed number that corresponds to the
maximum value allowed by the basis set used. Taking into account the
spin, the total occupancy for a given subspace $\Omega{}_{g}$ is
2, which is reflected in additional sum rules for the ONs, namely,
\begin{equation}
\sum_{p\in\Omega_{g}}n_{p}=1,\quad{\displaystyle 2\sum\limits _{p\in\Omega_{P}}n_{p}=2\sum_{g=1}^{N_{P}/2}}\sum_{p\in\Omega_{g}}n_{p}=N_{P}\label{sumNp}
\end{equation}
The simplest way to satisfy the constraints imposed on the two-particle
cumulant leads to PNOF5 \citep{Piris2011,Piris2013e}:
\begin{equation}
\begin{array}{c}
\Delta_{qp}=n_{p}^{2}\delta_{qp}+n_{q}n_{p}\left(1-\delta_{qp}\right)\delta_{q\Omega_{g}}\delta_{p\Omega_{g}}\\
\\
\Pi_{qp}=n_{p}\delta_{qp}+\Pi_{qp}^{g}\left(1-\delta_{qp}\right)\delta_{q\Omega_{g}}\delta_{p\Omega_{g}}\\
\\
\Pi_{qp}^{g}=\left\{ \begin{array}{cc}
-\sqrt{n_{q}n_{p}}\,, & p=g\textrm{ or }q=g\\
+\sqrt{n_{q}n_{p}}\,, & otherwise
\end{array}\right.\\
\\
\delta_{q\Omega_{g}}=\begin{cases}
1, & q\in\Omega_{g}\\
0, & q\notin\Omega_{g}
\end{cases};\quad g=1,2,\ldots,N_{P}/2\\
\\
\end{array}
\end{equation}
It is worth noting that $\Delta$ is a spin-independent matrix ($\Delta_{qp}=\Delta_{qp}^{\alpha\alpha}=\Delta_{qp}^{\alpha\beta}=\Delta_{qp}^{\beta\beta}$).
In addition, $\mathrm{\Delta_{\mathit{qp}}}$ and $\mathrm{\Pi_{\mathit{qp}}}$
are zero between orbitals belonging to different subspaces, so the
2-RDM reconstruction of PNOF5 corresponds to an independent pair model.
The resulting energy for the $N_{P}$ electrons is
\begin{equation}
\begin{array}{c}
E_{P}^{pnof5}=\sum\limits _{g=1}^{N_{P}/2}E_{g}+\sum\limits _{f\neq g}^{N_{P}/2}E_{fg}\\
\\
E_{g}=\sum\limits _{p\in\Omega_{g}}n_{p}\left(2\mathcal{H}_{pp}+\mathcal{J}_{pp}\right)+\sum\limits _{q,p\in\Omega_{g},q\neq p}\Pi_{qp}^{g}\mathcal{L}_{pq}\\
\\
E_{fg}=\sum\limits _{p\in\Omega_{f}}\sum\limits _{q\in\Omega_{g}}\left[n_{q}n_{p}\left(2\mathcal{J}_{pq}-\mathcal{K}_{pq}\right)\right]
\end{array}
\end{equation}
where $\mathcal{J}_{pp}=<pp|pp>$ is the Coulomb interaction between
two electrons with opposite spins at the spatial orbital $p$.

To go beyond the independent-pair approximation, $\mathrm{\Delta_{\mathit{qp}}=0}$
was kept, while nonzero $\Pi$-elements were considered between orbitals
belonging to different subspaces. Accordingly, new elements $\Pi_{qp}^{\Phi}=-\Phi_{q}\Phi_{p}$
with $\Phi_{q}=\sqrt{n_{q}(1-n_{q})}$ were included for orbitals
$q$ and $p$ belonging to different pairs (subspaces) in PNOF7 \citep{Piris2017,mitxelena2018a}.
Thus, the energy of the $N_{P}$ electrons becomes
\begin{equation}
E_{P}^{pnof7}=E_{P}^{pnof5}+\sum\limits _{f\neq g}^{N_{P}/2}\sum\limits _{p\in\Omega_{f}}\sum\limits _{q\in\Omega_{g}}\Pi_{qp}^{\Phi}\mathcal{L}_{pq}\label{interpair}
\end{equation}
On the other hand, taking into account Eq. (\ref{sumNp}), it follows
from (\ref{npmp}) that
\begin{equation}
\sum\limits _{p\in\Omega_{U}}m_{p}=2S=N_{U}
\end{equation}
To attain a high-spin multiplet state, let us take $m_{p}=1$ and
dim$\left\{ \Omega_{U}\right\} =N_{U}$. This assumption is trivial
for a doublet, but it is more restrictive for higher multiplets because
an underestimation of the energy can occur. Indeed, the correlation
$\Delta$ and $\Pi$ matrices, which determine the PNOF two-electron
cumulant, are null at the boundary values of the ONs \citep{Piris2013b}.
Therefore, the single-occupied orbitals ($n_{p}^{\alpha}=1,\;n_{p}^{\beta}=0,\;p\in\Omega_{U}$)
are not allowed, by construction, to contribute to the correlation
energy. This assumption is the most critical for spin-polarized systems.

After some algebra, the energy (\ref{PNOF}) for a spin-uncompensated
system can be written as
\begin{equation}
\begin{array}{c}
E^{pnofi}=E_{P}^{pnofi}+E_{PU}+E_{U}\quad(i=5,7)\qquad\\
\\
E_{PU}=\sum\limits _{p\in\Omega_{P}}\sum\limits _{q\in\Omega_{U}}n_{p}\left(2\mathcal{J}_{pq}-\mathcal{K}_{pq}\right)\qquad\\
\\
E_{U}=\sum\limits _{p\in\Omega_{U}}\mathcal{H}_{pp}+\frac{1}{2}\sum\limits _{p,q\in\Omega_{U}}\left(\mathcal{J}_{pq}-\mathcal{K}_{pq}\right)
\end{array}\label{E57u}
\end{equation}
\textcolor{black}{The solution is established by optimizing the energy
functional (}\ref{E57u}\textcolor{black}{) with respect to the ONs
and to the NOs, separately. The conjugate gradient method is used
for performing the optimization of the energy with respect to auxiliary
variables that enforce automatically the $N$-representability bounds
of the 1-RDM. In general, orbitals below }$N_{P}/2$\textcolor{black}{{}
are almost fully occupied, and those that are above the }single-occupied\textcolor{black}{{}
orbitals are almost empty, however, in a system with a strong non-dynamic
electron correlation, several NOs can have ONs equal or close to 1/2.}

\textcolor{black}{The self-consistent procedure proposed in \citep{Piris2009a}
yields the NOs by the iterative diagonalization of a Hermitian matrix
}\textbf{\textcolor{black}{F}}\textcolor{black}{. The off-diagonal
elements of }\textbf{\textcolor{black}{F}}\textcolor{black}{{} are determined
explicitly by the hermiticity of the Lagrange multipliers. The first-order
perturbation theory applying to each cycle of the diagonalization
process provides an aufbau principle for determining the diagonal
elements of }\textbf{\textcolor{black}{F}}\textcolor{black}{.}

In the following sections, we analyze the 1D Hubbard model with different
number of sites, and hydrogen chains with different ring sizes, for
different spin values, in order to test the NOFs given by \textcolor{black}{Eq.
(}\ref{E57u}\textcolor{black}{)} in strong non-dynamic correlation
regimes.

\section{\textcolor{black}{Hubbard model}}

\textcolor{black}{The Hubbard model is an ideal candidate for the
study of electron correlation in solid state physics and quantum chemistry.
In second quantization notation, the one-dimensional (1D) Hubbard
Hamiltonian takes the form \citep{Baeriswyl1995}
\begin{equation}
\begin{array}{c}
{\displaystyle \hat{H}=-t\sum_{\left\langle \mu,\upsilon\right\rangle ,\sigma}\left(\hat{a}_{\mu,\sigma}^{\dagger}\hat{a}_{\upsilon,\sigma}+\hat{a}_{\mu,\sigma}\hat{a}_{\upsilon,\sigma}^{\dagger}\right)+U\sum_{\mu}\hat{n}_{\mu,\alpha}\hat{n}_{\mu,\beta}}\end{array}
\end{equation}
where Greek indices $\mu$ and $\upsilon$ denote sites of the model,
$\left\langle \mu,\upsilon\right\rangle $ indicates only near-neighbors
interactions, $t$ is the hopping parameter, $U$ is the on-site inter-electron
repulsion parameter, and $\hat{n}_{\mu,\sigma}=\hat{a}_{\mu,\sigma}^{\dagger}\hat{a}_{\mu,\sigma}$
where $\hat{a}_{\mu,\sigma}^{\dagger}\left(\hat{a}_{\mu,\sigma}\right)$
corresponds to the fermionic creation(annihilation) operator. The
first term models the kinetic energy of the electrons hopping between
atoms, whereas the second term is the potential energy that emerges
from the on-site interaction $U$. }

\textcolor{black}{It is well-known that the Hartree-Fock (HF) approximation
retrieves the exact solution for the 1D Hubbard model at half-filling
if $U=0$, which in this limit corresponds to the tigh-binding model
\citep{Slater1954}. Conversely, in the $U/t\rightarrow\infty$ limit
the model becomes equivalent to the spin-1/2 Heisenberg model \citep{carten2015}.
Recently, the performance of commonly used NOFs for singlet states
in the 1D Hubbard model has been assessed \citep{Mitxelena2017,Mitxelena2018b,mitxelena2018a},
showing that PNOF7 is in good agreement with exact diagonalization
(ED) results for the Hubbard model at half-filling on even number
of sites. Since the purpose of this work is to introduce an approach
for studying spin-uncompensated systems with the NOFs developed in
our group, in this work we analyze the performance of spin-uncompensated
PNOFs by using the Hubbard model. Actually, our main goal is to study
many multiplicities in order to prove the advantages and shortcomings
of our new approximation.}

\textcolor{black}{Many spin multiplicities are of interest in electronic
structure theory. Thus, a complete test should include multiplicities
ranging from singlets to octets, the latter observed, for example,
in Gadolinium. In the following, we study the performance of spin-uncompensated
PNOF5 and PNOF7 by using the Hubbard model, and employing the $U/t$
ratio to cover all correlation regimes. The maximum number of sites
considered is 14 in order to compare with ED calculations. PNOF calculations
have been carried out using DoNOF code developed by M. Piris and coworkers
based on the iterative diagonalization method \citep{Piris2009a},
whereas ED results have been computed using a modified version of
the code developed by Knowles and Handy \citep{Knowles1984,Knowles1989}.}

\begin{figure}
\begin{centering}
\textcolor{black}{\includegraphics[scale=0.7]{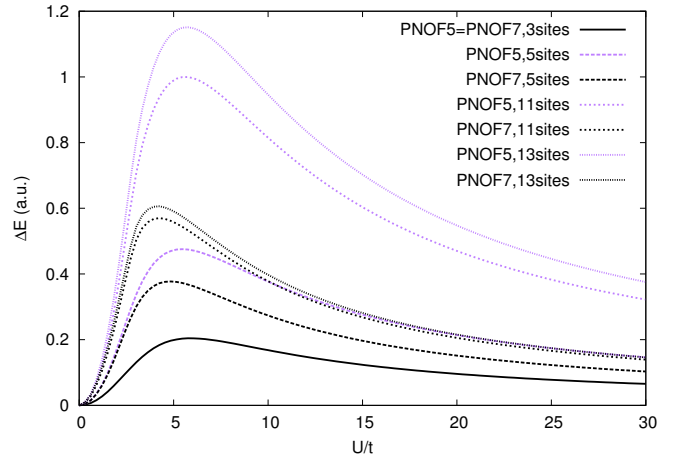}}
\par\end{centering}
\textcolor{black}{\caption{\textcolor{black}{Energy difference}~$\Delta E=$$E^{pnofi}-E^{ED}\:(i=5,7)$
for the \textcolor{black}{1D Hubbard model} with different number
of sites and $S=1/2$.\label{fig:1}}
}
\end{figure}

\begin{figure}
\begin{centering}
\textcolor{black}{\includegraphics[scale=0.7]{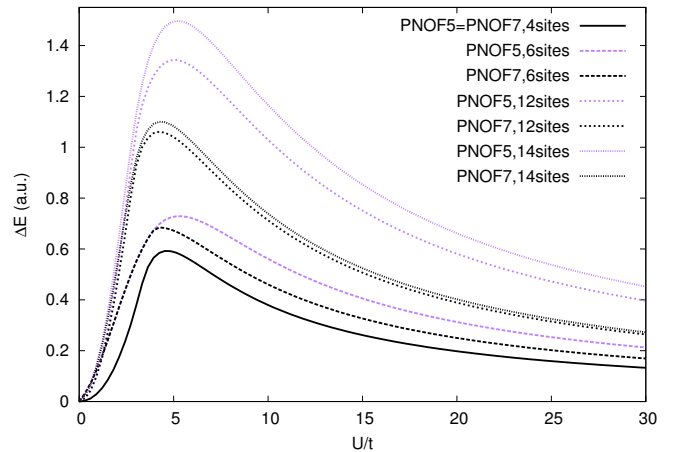}}
\par\end{centering}
\centering{}\textcolor{black}{\caption{\textcolor{black}{Energy difference}~$\Delta E=$$E^{pnofi}-E^{ED}\:(i=5,7)$
for \textcolor{black}{the 1D Hubbard model} with different number
of sites and $S=1$.\label{fig:2} }
}
\end{figure}

\begin{figure}
\begin{centering}
\textcolor{black}{\includegraphics[scale=0.7]{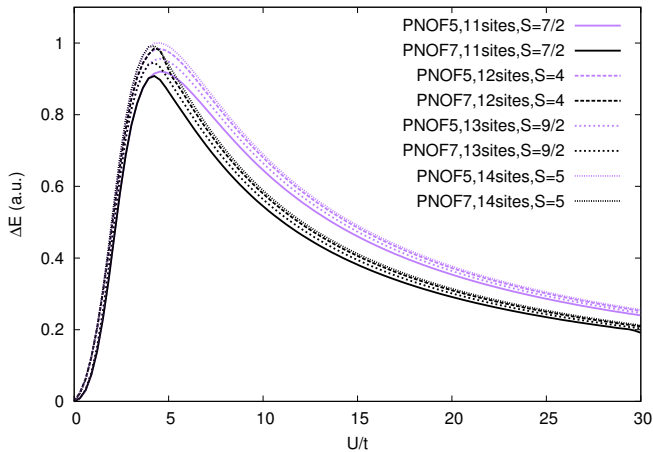}}
\par\end{centering}
\textcolor{black}{\caption{\label{fig:3}\textcolor{black}{{} Energy difference}~$\Delta E=$$E^{pnofi}-E^{ED}\:(i=5,7)$
for \textcolor{black}{the 1D Hubbard model} with different number
of sites and spins.}
}
\end{figure}

\textcolor{black}{In Figs. \ref{fig:1}-\ref{fig:3}, we show the
energy differences obtained with PNOF5 and PNOF7 with respect to the
ED values varying $U/t$. In general, we can observe that the error
increases to a maximum located at $U/t\thickapprox5$, and then seems
to decrease to a non-zero minimum at the $U/t\rightarrow\infty$ limit.
In comparison with the performance obtained for spin-compensated systems
\citep{Mitxelena2017,mitxelena2018a,Mitxelena2018b}, the errors shown
for spin-polarized systems are slightly larger for all correlation
regimes. Recall that PNOF5 takes into account the whole intra-pair
electron correlation, while PNOF7 also includes non-dynamic correlation
between electron pairs, so it is expected that errors with respect
to ED are greater for PNOF5 than for PNOF7.}

\textcolor{black}{At the $U/t\rightarrow\infty$ limit}, also known
as the strong correlation limit, electrons try to keep away one from
each other by half-filling the sites, which corresponds to the Mott-Hubbard
regime. In this case, there is no kinetic energy, all ONs are exactly
equal to $1/2$, which corresponds to a pure non-dynamic correlation
regime. Therefore, the fact that the \textcolor{black}{energy difference
in the $U/t\gg1$ region} is different from zero is related to the
lack of non-dynamic electron correlation. Even using PNOF7, our model
only takes into account part of this correlation between electrons
with opposite spins in the orbital subspace $\Omega_{P}$, that is,
the non-dynamic correlation of the $N_{P}$ paired electrons. As can
be seen from Eq. (\ref{E57u}), the $E_{PU}$ term is of HF type interactions
hence our approximate NOFs do not include correlation between electrons
with opposite spins involving the subspace $\Omega_{U}$. Singly occupied
orbitals do not contribute to the electron correlation (vide supra)\textcolor{black}{.}

\textcolor{black}{On the other hand, both functionals PNOF5 and PNOF7
underestimate dynamic inter-pair correlation effects, which turns
out to be important at low and intermediate $U/t$ values, namely
$U/t\approx5$. Recently \citep{Piris2017,Piris2018dyn}, one of us
(M.P.) has proposed a new method to recover the missing dynamic correlation
in spin-compensated systems by means of perturbation theory. Corrections
of this type are not within the scope of this work and will be incorporated
for spin-polarized systems in the future. In what follows, we focus
mainly on the region where the effects of non-dynamic correlation
predominate, that is, for large $U/t$.}

\textcolor{black}{}

\textcolor{black}{In Fig. \ref{fig:1}, we plot the results for systems
with odd number of sites (3, 5, 11 and 13) and spin $S=1/2$ (}$N_{U}=1$\textcolor{black}{).
Note that there are no differences between PNOF5 and PNOF7 in the
3 sites system with $S=1/2$ because there is only one electron pair
with opposite spins considered (}$N_{P}=2$\textcolor{black}{). }We
also note that errors with respect to ED results increase by rising
sites from 5 \textcolor{black}{(}$N_{P}=4$\textcolor{black}{)} to
13 \textcolor{black}{(}$N_{P}=12$\textcolor{black}{)}. Indeed, the
electron correlation is increasingly ignored by accounting for more
$N_{P}$ pairs for one unpaired electron. \textcolor{black}{The introduction
in Eq. (\ref{interpair}) of the term to account for non-dynamic inter-pair
correlation makes PNOF7 more robust with the increase in the number
of electron pairs, which is consistent with the results obtained for
singlet states \citep{mitxelena2018a}.}

\textcolor{black}{}

\textcolor{black}{In Fig. \ref{fig:2}, the triplet states (}$N_{U}=2$\textcolor{black}{)
are considered for systems with an even number of sites. It is worth
mentioning that for 4 sites, the PNOF5 and PNOF7 values coincide due
to having only one pair with opposite spins (}$N_{P}=2$\textcolor{black}{).
We observe that the trend is similar to that obtained for $S=1/2$,
however, the errors are higher in the $S=1$ case (notice that the
scale of the $\Delta$}E\textcolor{black}{{} axis is different between
figures \ref{fig:1} and \ref{fig:2}). There are two unpaired electrons,
whose correlation effects are neglected, so that errors increase for
all systems with $S=1$ with respect to the corresponding systems
with $S=1/2$.}

Figures 1 and 2 show how our approximation behaves for a given $N_{U}$
by varying the value of $N_{P}$. In both figures, we observe that
the curves approximate each other by increasing the number of sites
for a given value of \textcolor{black}{$U/t$}, especially in the
region of high non-dynamic correlation. Consequently, the error with
respect to the ED values tends to stabilize.

\textcolor{black}{}

We are also interested in the performance of our approach for a fixed
$N_{P}$. Consequently, we decided to analyze which effects cause
the increase of the system size and its spin multiplicity at the same
time. In Fig. \textcolor{black}{\ref{fig:3}, we start with the 11
sites Hubbard model (}$N_{U}=7,\,S=7/2$\textcolor{black}{) with only
two electron pairs (}$N_{P}=4$\textcolor{black}{). Then, we keep
fixed }$N_{P}=4$\textcolor{black}{{} and add an additional site and
an extra unpaired electron consecutively (}$N_{U}=8,9,10,\,S=4,9/2,5$\textcolor{black}{).
Neither PNOF5 nor PNOF7 yield significant differences for $U/t<5$,
however, for larger $U/t$ values, where non-dynamic correlation regimes
prevail, PNOF7 performs better than PNOF5 as expected. This result
indicates that the error of our functionals with respect to the ED
results seems to stabilize for the variation of the spin multiplicity
together with the size for a given number} $N_{P}$\textcolor{black}{{}
of electron pairs as well.}

\textcolor{black}{}
\begin{figure}
\begin{centering}
\textcolor{black}{\includegraphics[scale=0.7]{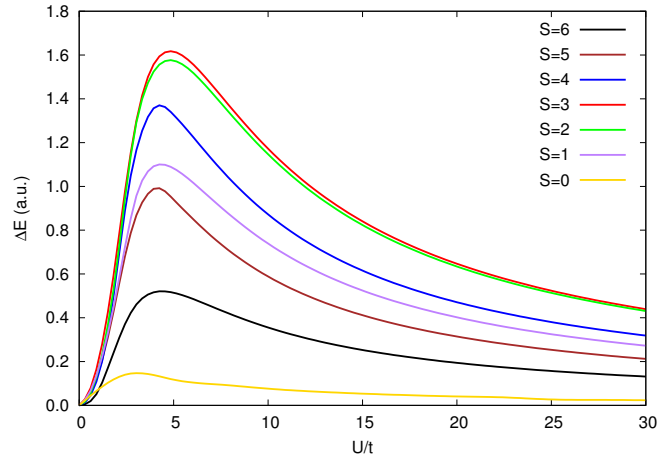}}
\par\end{centering}
\textcolor{black}{\caption{\textcolor{black}{Energy difference}~$\Delta E=$$E^{pnof7}-E^{ED}$
for \textcolor{black}{the 1D Hubbard model} with 14 sites and different
spin values.\label{fig:4}}
}
\end{figure}

In Fig. \textcolor{black}{\ref{fig:4}, we have collected the results
obtained with PNOF7 for different multiplicities and a given number
of electrons (14 sites Hubbard model). The best results are obtained
for minimum and maximum multiplicities. For $S=0$, a great amount
of electron correlation is recovered by PNOF7 for any correlation
regime.} For large spin values, the HF energy components dominate
up to the full polarization (S = 7) in which there is no correlation
energy. For the latter, the exact result is reduced to the HF energy,
which is also recovered with our approach. For the intermediate spin
values, the interactions related to the \textcolor{black}{$E_{PU}$}
energy component in Eq. \textcolor{black}{(\ref{E57u})} should be
improved to reduce the relatively large errors in Fig. \textcolor{black}{\ref{fig:4}}.

\section{\textcolor{black}{Hydrogen rings}}

\textcolor{black}{The lack of long-range inter-electronic interactions
may be the most important limitation of the Hubbard model. In this
section, we focus on rings composed of hydrogen atoms, which model
the non-dynamic electronic correlation in the presence of long-range
interaction effects, to examine whether the results obtained in the
previous section are still valid. Here, }$<kl|ij>$ in Eq. (\ref{NOF})
are the matrix elements of the bare Coulomb interaction.\textcolor{black}{{}
These systems have previously been used to perform benchmarking of
many-body approximations \citep{Boguslawski2014,Mitxelena2017}. Theoretical
investigations on thermodynamic stabilities \citep{yamaguchi1983,Wright1992}
and aromaticity \citep{Jiao1996} of hydrogen rings have also been
carried out.}

\textcolor{black}{}
\begin{figure}
\begin{centering}
\textcolor{black}{\includegraphics[scale=0.65]{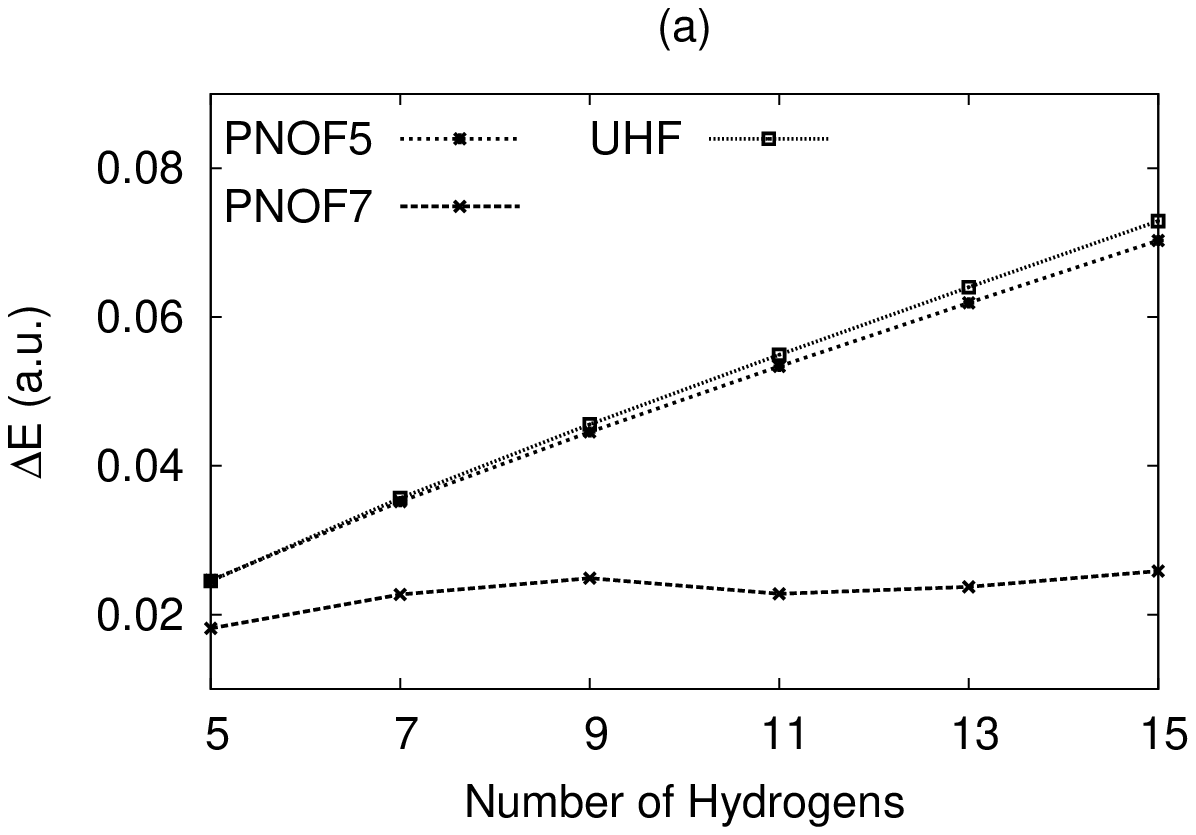}}
\par\end{centering}
\begin{centering}
\textcolor{black}{\includegraphics[scale=0.65]{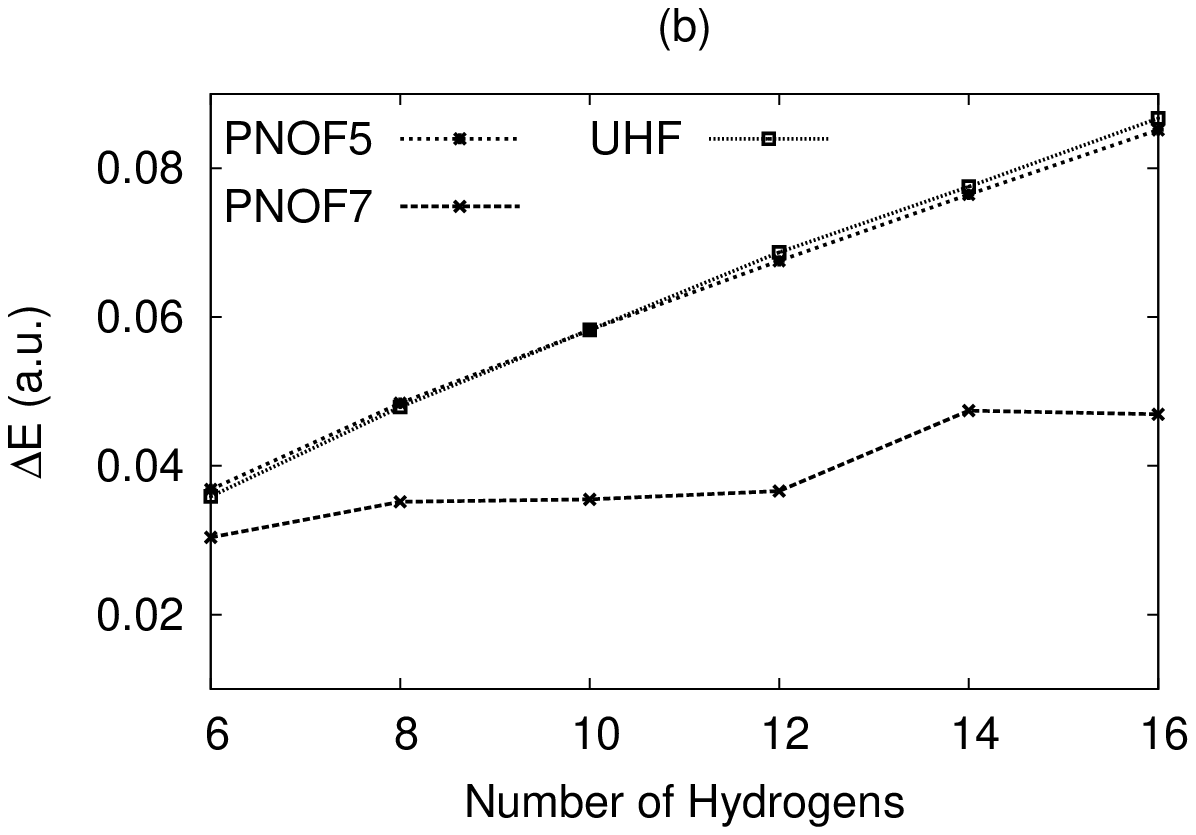}}
\par\end{centering}
\begin{centering}
\textcolor{black}{\includegraphics[scale=0.65]{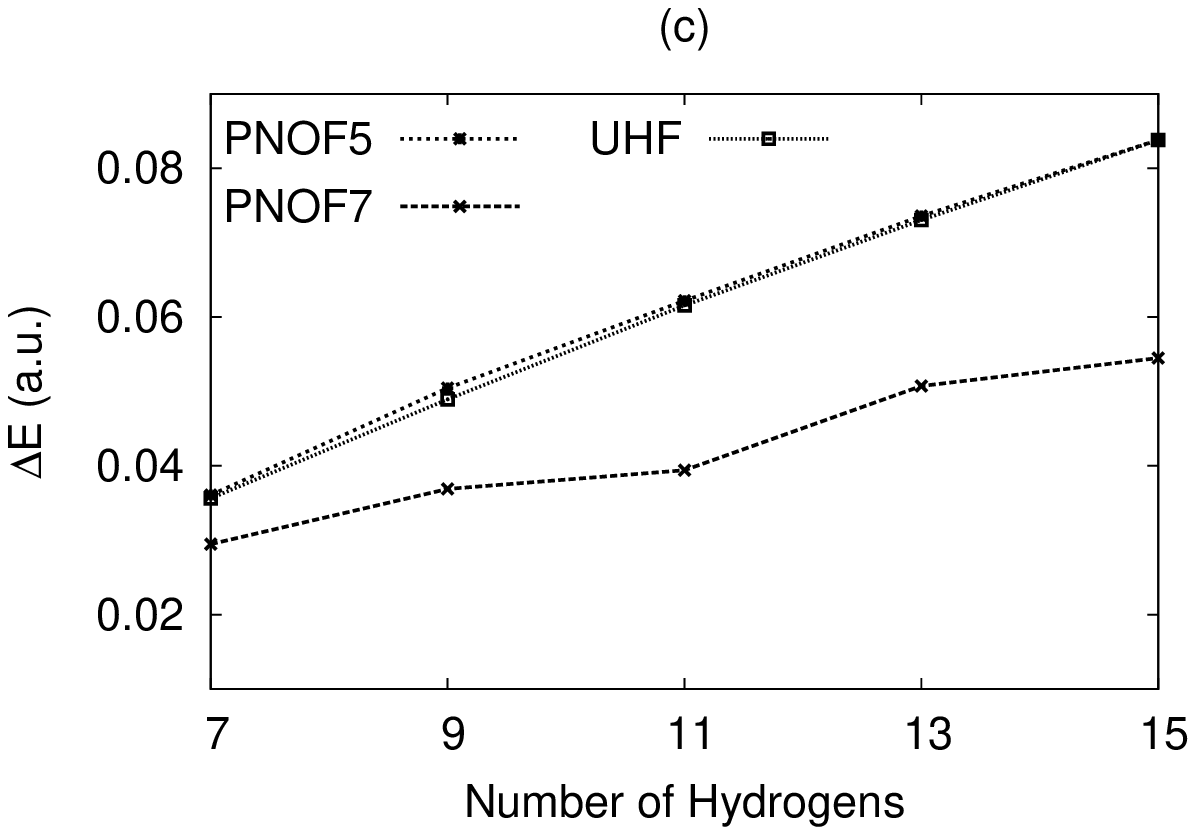}}
\par\end{centering}
\begin{centering}
\textcolor{black}{\includegraphics[scale=0.65]{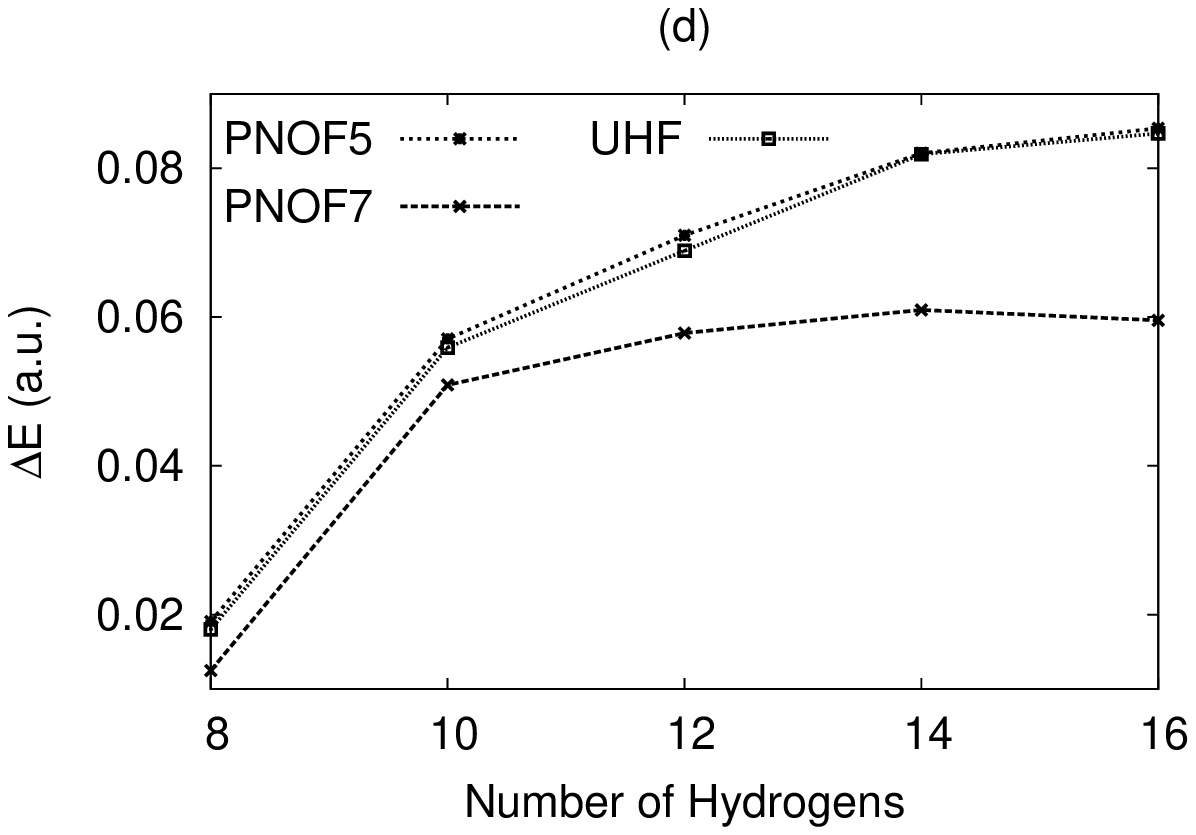}}
\par\end{centering}
\textcolor{black}{\caption{\label{fig:5} Energy differences $\Delta E=$$E^{pnofi}-E^{ED}\:(i=5,7)$
as a function of the number of hydrogens obtained with the STO-3G
basis set. From top to down, $S=1/2(a),1(b),3/2(c)$ and $2(d)$,
respectively.}
}
\end{figure}
\textcolor{black}{}
\begin{figure}
\begin{centering}
\textcolor{black}{\includegraphics[scale=0.65]{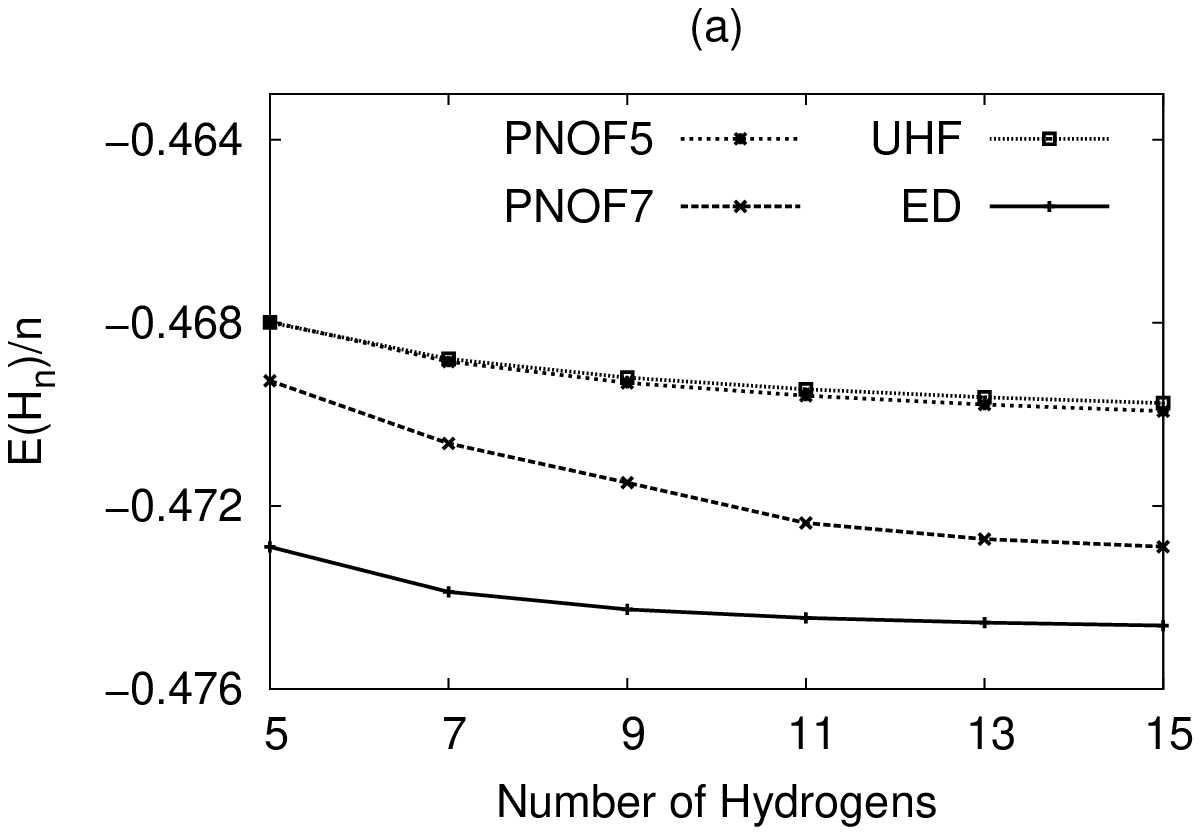}}
\par\end{centering}
\begin{centering}
\textcolor{black}{\includegraphics[scale=0.65]{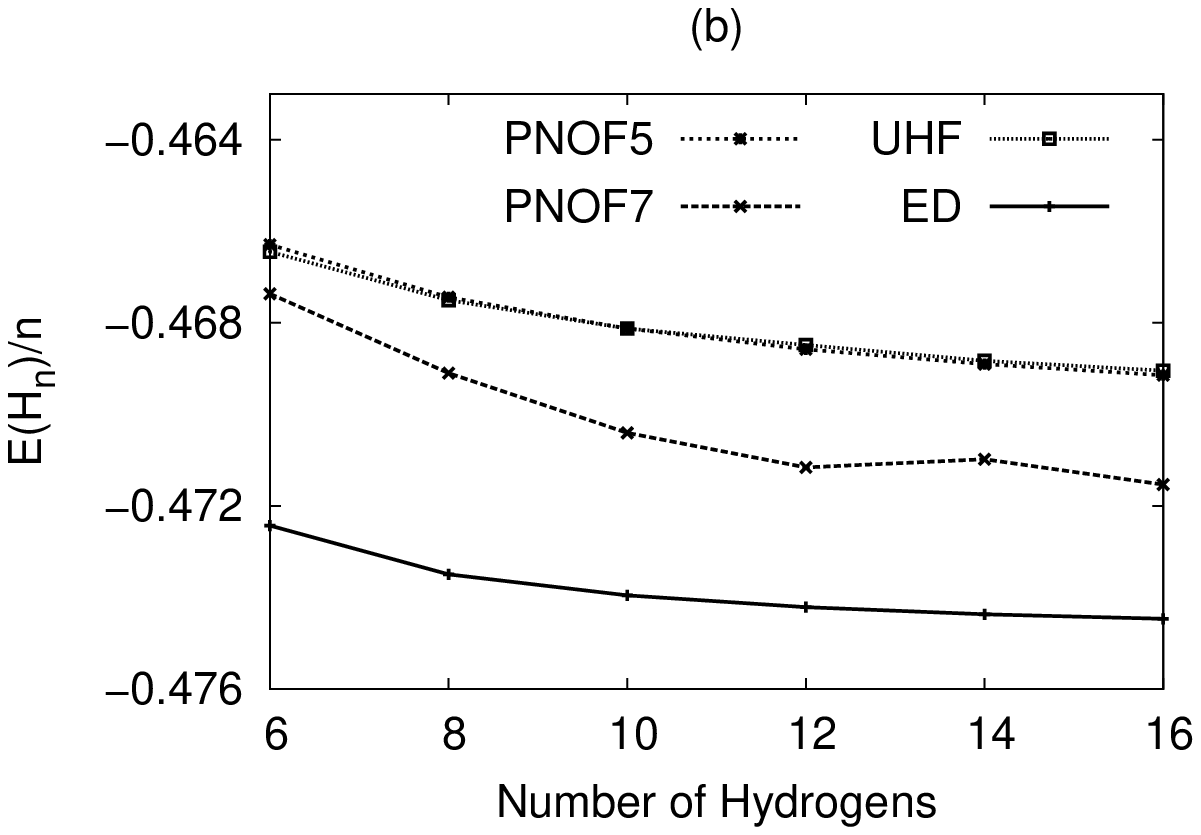}}
\par\end{centering}
\begin{centering}
\textcolor{black}{\includegraphics[scale=0.65]{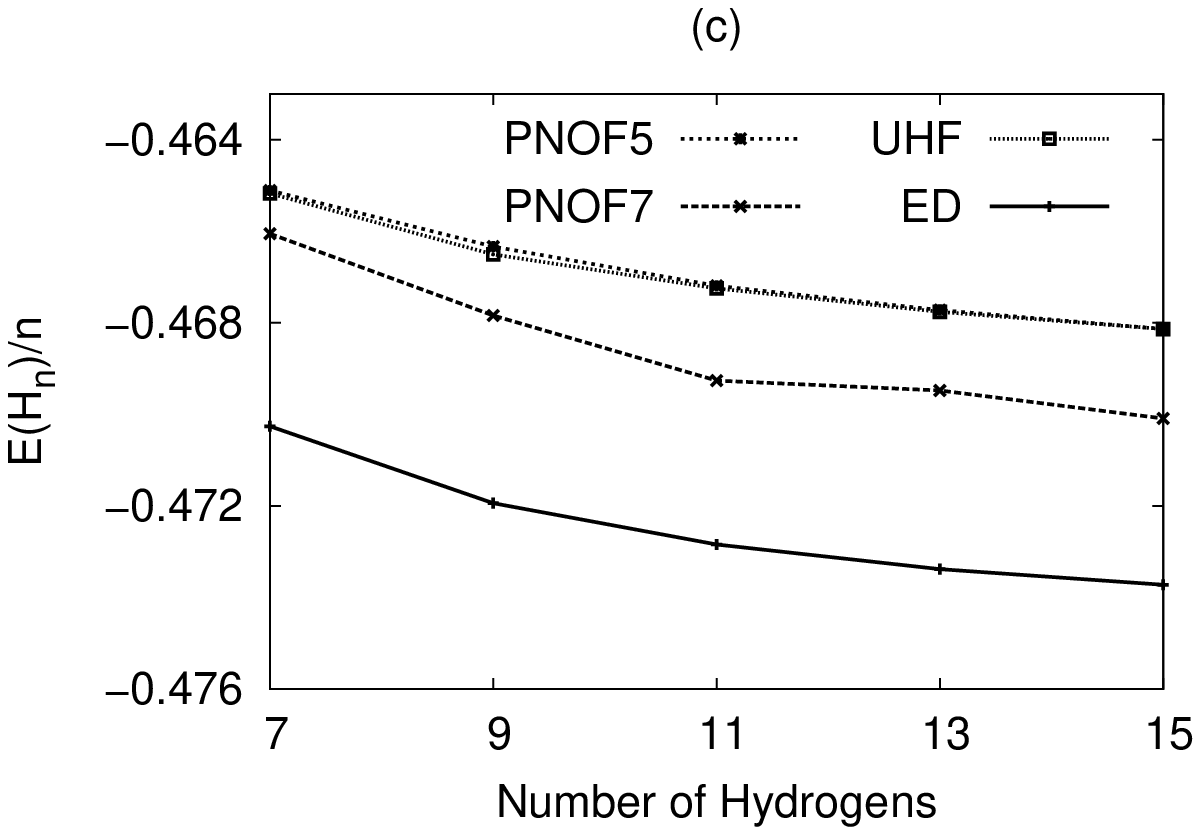}}
\par\end{centering}
\begin{centering}
\textcolor{black}{\includegraphics[scale=0.65]{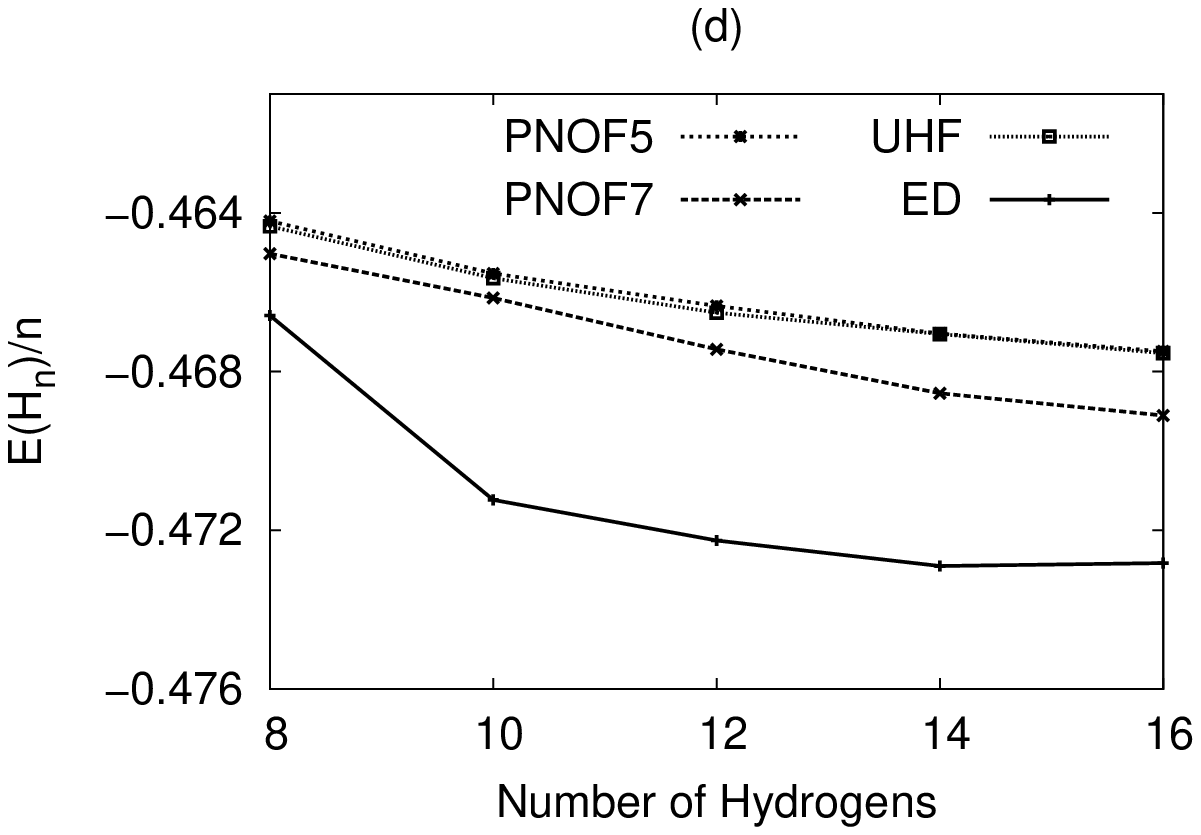}}
\par\end{centering}
\textcolor{black}{\caption{\label{fig:6} The unit cell energy ($E(H_{n})/n$) as a function
of the number of hydrogens ($n$) obtained with the STO-3G basis set.
From top to down, $S=1/2(a),1(b),3/2(c)$ and $2(d)$, respectively.}
}
\end{figure}

\textcolor{black}{In Fig. \ref{fig:5}, we plot total energy differences
with respect to the ED results obtained by using PNOF5, PNOF7 and
unrestricted HF (UHF) for} $S=1/2(a),1(b),3/2(c)$ and $2(d)$\textcolor{black}{.
The maximum number of hydrogens considered was 16 in order to compare
with ED calculations. Total energies corresponding to ED and UHF have
been carried out by means of the Psi4 suite of programs \citep{Parrish2017}.
All calculations were done with the minimal STO-3G basis set \citep{sto-3g}.
The inter-nuclear distance between hydrogen atoms was fixed to $R_{H-H}=2.0\:\textrm{Å}$
in order to have mostly non-dynamic electron correlation \citep{Mazziotti-hydrogen-chain}.
Interestingly, PNOF5 and UHF behave similarly for any spin multiplicity,
however, UHF retrieves electron correlation }at the expense of breaking
the \textcolor{black}{total spin symmetry, while PNOF5 affords the
correct value of $<\hat{S}^{2}>$.}

\textcolor{black}{As already seen for the Hubbard model, the error
obtained when using PNOF7 is significantly reduced with respect to
PNOF5.} Taking into account the curves shown in Fig. \textcolor{black}{\ref{fig:5},
}we can observe that \textcolor{black}{the difference $E^{pnof7}-E^{ED}$
}seems to tend earlier to a constant value than the energies of PNOF5
and UHF, which produces larger errors as the number of hydrogens increases,
reaching the asymptote presumably for much larger rings. \textcolor{black}{This
represents an important step forward in the description of spin-polarized
systems with strong correlation effects, since PNOF7 shows a significant
improvement over the widely used mean-field UHF method.}

Finally, in Fig. \ref{fig:6} we report the total energies ($E(H_{n})$)
divided by the number ($n$) of hydrogens as a function of the ring
size using PNOF5, PNOF7, UHF and ED\textcolor{black}{. }Comparing
$E(H_{n})/n$ obtained with the different approximations \textcolor{black}{with
respect to the ED }values, we observe that PNOF7 is in good agreement
with the latter, although the error increases slightly as S increases.

\section{\textcolor{black}{Closing Remarks}}

The extension of PNOF5 and PNOF7 has been achieved for systems with
spin polarization. The obtained natural orbital functionals provide
an adequate value of\textcolor{black}{{} $<\hat{S}^{2}>$ and $<\hat{S}_{z}>$,}
so that the spin contamination effects are not present. As in the
case of singlet states, \textcolor{black}{N-representability conditions
}on the reconstructed two-particle reduced density matrix are satisfied
for high-spin multiplet state.

In the approach proposed here, we have considered a subspace of fully-singly
occupied orbitals with the same spin, while the rest of the electron
pairs with opposite spins are distributed in the remaining orbital
subspace. In this way, the electron pairs form a singlet, and the
unpaired electrons are responsible for the spin of the system. This
simplification works well for a doublet, but it is more restrictive
for higher multiplets because the single-occupied orbitals are not
allowed to contribute to the correlation energy.

Despite its simplicity, the model presented here is able to provide
a qualitatively correct description of the strong-correlation effects
that appear in the one-dimensional Hubbard model and the hydrogen
rings. A behavior similar to that obtained for singlets, especially
for doublets and triplets, was achieved for these systems, so it has
been shown that the performance of PNOF5 and PNOF7 does not deteriorate
when they are extended to spin-polarized systems.

\textcolor{black}{PNOF5 fully takes into account the intra-pair electron
correlation, while PNOF7 also includes non-dynamic correlation between
electron pairs, so the errors with respect to exact diagonalization
are greater for PNOF5 than for PNOF7.} It was shown that
errors seem to stabilize as the system size increases regardless of
the value of the spin. In addition, PNOF7 showed a significant improvement
over the widely used unrestricted Hartree-Fock method for spin-polarized
systems. \textcolor{black}{Therefore, PNOF7 is an ideal candidate
to be employed in the study of spin-uncompensated systems composed
of a large number of atoms.}

\textcolor{black}{The best results were obtained for minimum and maximum
multiplicities.} For the intermediate spin values, the interactions
related to the \textcolor{black}{paired-unpaired} energy component
should be improved to reduce the relatively errors. \textcolor{black}{In
this vein, an explicit form of the two-electron cumulant of the reconstructed
two-particle reduced matrix capable to describe fractional occupancies
of the unpaired electrons is still missing. A work in this direction
is underway.}

\section{\textcolor{black}{Acknowledgments}}

\selectlanguage{american}%
Financial support comes from Ministerio de Economía y Competitividad
(Ref. CTQ2015-67608-P). The authors thank for technical and human
support provided by IZO-SGI SGIker of UPV/EHU and European funding
(ERDF and ESF). R. Q-M. is grateful to\foreignlanguage{english}{ Conacyt
for the grant ``Fronteras'' (Project 867) and the PhD grant 587950.
I. M. }is grateful to Vice-Rectory for research of the UPV/EHU for
the PhD. grant\foreignlanguage{english}{ (PIF//15/043).}

\selectlanguage{english}%

\end{document}